\RequirePackage{ifpdf}
\ifpdf % We are running pdfTeX in pdf mode
\documentclass[pdftex]{sigma}
\else
\documentclass{sigma}
\fi

\begin{document}

\renewcommand{\PaperNumber}{088}

\FirstPageHeading

\renewcommand{\thefootnote}{$\star$}

\ShortArticleName{Complex Projection of Quasianti-Hermitian
Quaternionic Hamiltonian Dynamics}

\ArticleName{Complex Projection of Quasianti-Hermitian\\
Quaternionic Hamiltonian Dynamics\footnote{This paper is a
contribution to the Proceedings of the 3-rd Microconference
``Analytic and Algebraic Me\-thods~III''. The full collection is
available at
\href{http://www.emis.de/journals/SIGMA/Prague2007.html}{http://www.emis.de/journals/SIGMA/Prague2007.html}}}

\Author{Giuseppe SCOLARICI}

\AuthorNameForHeading{G. Scolarici}

\Address{Dipartimento di Fisica dell'Universit\`{a} del Salento,
and INFN,\\ Sezione di Lecce, I-73100 Lecce, Italy}
\Email{\href{mailto:scolarici@le.infn.it}{scolarici@le.infn.it}}

\ArticleDates{Received July 05, 2007, in f\/inal form September
03, 2007; Published online September 08, 2007}

\Abstract{We characterize the subclass of quasianti-Hermitian
quaternionic Hamiltonian dynamics such that their complex
projections are one-parameter semigroup dynamics in the space of
complex quasi-Hermitian density matrices. As an example, the
complex projection of a spin-$\frac{1}{2}$ system in a constant
quasianti-Hermitian quaternionic potential is considered.}


\Classification{81P68; 15A33}

\section{Introduction}

Many years ago, it was shown, by using some lattice theoretic
arguments, that it is possible to consider the set of states of a
quantum system as a vector space over the real, complex or
quaternionic f\/ields \cite{bir}. However, the research on
quaternionic quantum mechanics (QQM) began much later \cite{fin}
and pursued up to now. A systematic study of QQM is given
in~\cite{Adler}. In this context, some recent studies have been
developed along two seemingly uncorrelated lines.

On one hand, after a seminal idea by Kossakowski \cite{Kossa1},
the complex projection of dyna\-mics generated by (time independent)
quaternionic anti-Hermitian Hamiltonians was studied by showing
that they represent one-parameter semigroup dynamics in the space
of complex (Hermitian) density matrices
\cite{compent,Asor,gallipoli,ASS}. This peculiarity can be useful,
for instance, in the case of two qubits compound system, the
complex projection of quaternionic unitary dynamics bet\-ween pure
states permits description of interesting phenomena as decoherence
and optimal entanglement generation~\cite{ASS}.

On the other hand, motivated by the recent studies on
non-Hermitian Hamiltonians~\cite{proc}, pseudoanti-Hermitian
quaternionic Hamiltonians were introduced in order to generalize
standard anti-Hermitian Hamiltonians in quaternionic Hilbert
space~\cite{scola}. The dynamics generated by such (time
independent) Hamiltonians preserves an alternative (in general)
indef\/inite inner product in the Hilbert space. In this context,
it was proven that if (and only if) the pseudoanti-Hermitian
quaternionic Hamiltonians belong to the subclass of
quasianti-Hermitian ones, then, an alternative positive def\/inite
quaternionic inner product preserved by the corresponding dynamics
can be introduced~\cite{altern}.

Moreover, it was shown that the complex quasi-Hermitian systems
can be described as open quantum systems. Indeed, a master
equation of Lindblad--Kossakowski type can be derived for such
systems, obtaining one-parameter semigroup dynamics in the space
of complex quasi-Hermitian density matrices~\cite{bologna}.

Motivated by these three apparently separated topics, we intend to
study, in this paper, the complex projection of
quasianti-Hermitian quaternionic Hamiltonian dynamics. We will
show that if (and only if) the Hamiltonian is $\eta
$-quasianti-Hermitian with respect a (Hermitian positive
def\/inite) \textit{complex} $\eta $, then the corresponding
complex projection dynamics generates one-parameter semigroup
dynamics in the space of complex quasi-Hermitian density matrices.
This result, allows us to produce systematically, via the complex
projection operation, one-parameter semigroup dynamics in the
space of complex quasi-Hermitian density matrices. Moreover, we
will obtain the complex projection of quaternionic anti-Hermitian
Hamiltonian dynamics as a~very particular case of this more
general setting.

The plan of the paper is the following: In Section~2 we will
introduce the density matrix formalism for quaternionic spaces and
discuss the complex projection of anti-Hermitian quaternionic
Hamiltonian dynamics. In Section~3 the notion of quaternionic
pseudo-Hermitian density matrix is introduced and a corresponding
Liouville-von Neumann type equation is derived. Moreover, the
complex projection of quasianti-Hermitian quaternionic dynamics is
considered in Section~4, where the subclass of quasianti-Hermitian
quaternionic Hamiltonian dynamics such that their complex
projections are one-parameter semigroup dynamics in the space of
complex quasi-Hermitian density matrices is characterized. In
order to illustrate these results, the complex projection of a
spin-$\frac{1}{2}$ system in a constant quasianti-Hermitian
quaternionic potential is studied in the last section.

\section{Complex projection of anti-Hermitian quaternionic\\ Hamiltonian
dynamics}

We will discuss, in this section, some results about the complex
projection of anti-Hermitian quaternionic Hamiltonian dynamics.

A (real) quaternion is usually expressed as
\begin{gather*}
q=q_{0}+q_{1}i+q_{2}j+q_{3}k,
\end{gather*}%
where $q_{l}\in \mathbb{R}$ $\left( l=0,1,2,3\right) $,
$i^{2}=j^{2}=k^{2}=-1$, $ij=-ji=k$.

The quaternion skew-f\/ield $\mathbb{Q}$ is an algebra of rank $4$ over $%
\mathbb{R}$, non commutative and endowed with an involutive
anti-automorphism (\textit{conjugation}) such that
\begin{gather*}
q\rightarrow \bar{q}=q_{0}-q_{1}i-q_{2}j-q_{3}k.
\end{gather*}

The density matrix $\rho _{\psi }$ associated with a pure state
$|\psi
\rangle $ belonging to a quaternionic $n$-di\-mensional right Hilbert space $%
\mathbb{Q}^{n}$ is def\/ined by \cite{Adler}
\begin{gather*}
\rho _{\psi }=|\psi \rangle \langle \psi |  %\label{ro}
\end{gather*}
and is the same for all normalized ray representatives. By
def\/inition, density matrices $\rho _{\psi }$ associated with
pure states, are represented by rank one, positive def\/inite
quaternionic Hermitian operators on $\mathbb{Q}^{n}$ with unit
trace. In analogy with standard quantum mechanics (CQM),
quaternionic mixed states are described by positive quaternionic
Hermitian operators (density matrices) $\rho $ on $\mathbb{Q}^{n}$
with unit trace and rank greater than one.

The expectation value of a quaternionic Hermitian operator $O$ on a state $%
|\psi \rangle $ can be expressed in terms of $\rho _{\psi }$ as
\cite{Adler}
\begin{gather*}
\langle O\rangle _{\psi }=\langle \psi |O|\psi \rangle ={\rm Re}\,\mathrm{Tr}%
(O|\psi \rangle \langle \psi |)={\rm Re}\,\mathrm{Tr}(O\rho _{\psi
}).
%\label{expec}
\end{gather*}

Expanding $O=O_{\alpha }+jO_{\beta }$ and $\rho =\rho _{\alpha
}+j\rho _{\beta }$ \ in terms of complex matrices $O_{\alpha }$,
$O_{\beta }$, $\rho
_{\alpha }$ and~$\rho _{\beta }$, it follows that the expectation value $%
\langle O\rangle _{\psi }$ may depend on $O_{\beta }$ or $\rho
_{\beta }$ only if both $O_{\beta }$ and $\rho _{\beta }$ are
dif\/ferent from zero. Indeed,
\begin{gather*}
\langle O\rangle ={\rm Re}\,\mathrm{Tr}(O\rho )={\rm Re}\,\mathrm{Tr}%
(O_{\alpha }\rho _{\alpha }-O_{\beta }^{\ast }\rho _{\beta }),
%\label{esplexpc}
\end{gather*}%
where $\ast $ denotes complex conjugation.

Thus, the expectation value of an Hermitian operator $O$ on the
state $\rho $ depends on the quaternionic parts of $O$ and $\rho
$, only if both the observable and the state are represented by
genuine quaternionic matrices.

However, if an observable $O_{\alpha }$ is described by a pure
\emph{complex} Hermitian matrix, its expectation value does not
depend on the quaternionic part $j\rho _{\beta }$\ of the state
$\rho =\rho _{\alpha }+j\rho _{\beta }$.

Now, let us denote by $M(\mathbb{Q})$ and $M(\mathbb{C})$ the space of $%
n\times m$ quaternionic and complex matrices respectively and let $%
M=M_{\alpha }+jM_{\beta }\in M(\mathbb{Q})$. We def\/ine the
complex projection
\begin{gather*}
P:M(\mathbb{Q})\rightarrow M(\mathbb{C})
\end{gather*}%
by the relation
\begin{gather}
P[M]=\tfrac{1}{2}[M-iMi]=M_{\alpha }.  \label{complexProjection}
\end{gather}

By recalling that the complex projection of a quaternionic density
matrix is a complex density matrix \cite{Asor}, we can conclude
that\emph{\ the expectation value predicted in the standard CQM
for the state }$\rho
_{\alpha }$ \emph{coincides with the one predicted in QQM for the state }$%
\rho $\emph{,} since
\begin{gather*}
\mathrm{Tr}(O_{\alpha }\rho _{\alpha })={\rm
Re}\,\mathrm{Tr}(O_{\alpha }\rho _{\alpha })={\rm
Re}\,\mathrm{Tr}(O_{\alpha }\rho ).
\end{gather*}%
This simple observation is actually very important in our
approach, in that it enables us to merge CQM in the (more general)
framework of QQM, without modifying any theoretical prediction (as
long as complex observables are taken into account).

Moreover, when we consider quaternionic unitary dynamics,
\begin{gather}
\rho (t)=U(t)\rho (0)U^{\dagger }(t),  \label{unitev}
\end{gather}%
where
\begin{gather*}
U(t)=e^{-Ht}=U_{\alpha }+jU_{\beta }  %\label{timeordev}
\end{gather*}%
with $H=H_{\alpha }+jH_{\beta }=-H^{\dagger }$, the dif\/ferential
equation associated with the time evolution for $\rho $ reads
\begin{gather}
\frac{d}{dt}\rho (t)=-[H,\rho (t)].  \label{eveqtimdep}
\end{gather}%
In addition, equations (\ref{unitev}) and (\ref{eveqtimdep})
reduce to
\begin{gather*}
\rho _{\alpha }(t)=U_{\alpha }\rho _{\alpha }(0)U_{\alpha
}^{\dagger }+U_{\beta }^{\ast }\rho _{\alpha }^{\ast }(0)U_{\beta
}^{T}+U_{\alpha }\rho _{\beta }^{\ast }(0)U_{\beta }^{T}-U_{\beta
}^{\ast }\rho _{\beta
}(0)U_{\alpha }^{\dagger }  %\label{finitecomppro}
\end{gather*}%
and
\begin{gather}
\frac{d}{dt}\rho _{\alpha }=-[H_{\alpha },\rho _{\alpha
}]+H_{\beta }^{\ast }\rho _{\beta }-\rho _{\beta }^{\ast }H_{\beta
},  \label{gencompro}
\end{gather}%
respectively, for the complex projection of the density
matrix~\cite{Asor}.

It was proven that the dynamics ruled by
equation~(\ref{gencompro}) is a one-parameter semigroup dynamics
in the space of complex density matrices~\cite{Asor}.

\section{Pseudo-Hermitian quaternionic density matrices\\ and their dynamics}

In this section, we will introduce the notion of quaternionic
pseudo-Hermitian density matrix and a corresponding Liouville--von
Neumann type equation will be derived.

Denoting by $Q^{\ddagger }$ the adjoint of an operator $Q$ with
respect to the pseudo-inner product
\begin{gather}
(\cdot ,\cdot )_{\eta }=(\cdot ,\eta \cdot )  \label{IP}
\end{gather}%
(where $(\cdot ,\cdot )$ represent the standard quaternionic inner
product
in the space $\mathbb{Q}^{n}$), we have%
\begin{gather}
Q^{\ddagger }=\eta ^{-1}Q^{\dagger }\eta  \label{adj}
\end{gather}%
so that for any $\eta $-pseudo-Hermitian operator, i.e.,
satisfying the relation,
\begin{gather}
\eta Q\eta ^{-1}=Q^{\dagger },  \label{pseudoh}
\end{gather}%
one has, $Q=Q^{\ddagger }$. These operators constitute the
physical observables of the system~\cite{obse}.

If $Q$ is $\eta $-pseudo-Hermitian, equation~(\ref{pseudoh})
immediately implies that $\eta Q$ is Hermitian, so that the
expectation value of $Q$ in the state $|\psi \rangle $ with
respect to the pseudo-inner product (\ref{IP}) can be obtained,
\begin{gather}
\left\langle \psi \right| \eta Q\left| \psi \right\rangle ={\rm Re}\,\mathrm{%
Tr}(|\psi \rangle \langle \psi |\eta Q)={\rm
Re}\,\mathrm{Tr}(\tilde{\rho}Q), \label{ex}
\end{gather}%
where $\tilde{\rho}=|\psi \rangle \langle \psi |\eta $.

More generally, if $\rho $ denotes a generic quaternionic
(Hermitian,
positive def\/inite) density matrix, we can associate it with a \textit{%
generalized} density matrix $\tilde{\rho}$ by means of a
one-to-one mapping in the following way:
\begin{gather}
\tilde{\rho}=\rho \eta  \label{rho}
\end{gather}%
and obtain $\left\langle Q\right\rangle _{\eta }={\rm Re}\,\mathrm{Tr}(\tilde{%
\rho}Q).$

Note that $\tilde{\rho}$ is $\eta $-pseudo-Hermitian:
\begin{gather*}
\tilde{\rho}^{\dagger }=\eta \rho =\eta \tilde{\rho}\eta ^{-1}.
\end{gather*}

As in the Hermitian case discussed in the previous section,
equation (\ref{ex}) immediately implies that the expectation value
of an $\eta $-pseudo-Hermitian operator $Q$ on the generalized
state $\tilde{\rho}$ depends on the quaternionic parts of $Q$ and
$\tilde{\rho}$, only if both the operator and the generalized
state are represented by genuine quaternionic matrices. Hence, if
a $\eta $-pseudo-Hermitian operator~$Q$ is described by a
\emph{complex} matrix, its expectation value does not depend
on the quaternionic part~$j\tilde{\rho}_{\beta }$\ of the generalized state $%
\tilde{\rho}=\tilde{\rho}_{\alpha }+j\tilde{\rho}_{\beta }$.

It was shown that whenever the quaternionic Hamiltonian $H$ of a
quantum system is $\eta $-pseudoanti-Hermitian, i.e.,
\begin{gather*}
\eta H\eta ^{-1}=-H^{\dagger }, % \label{pseudoermiticity}
\end{gather*}%
where $\eta =\eta ^{\dagger }$, the pseudo-inner product
(\ref{IP}), is invariant under the time translation generated
by~$H$~\cite{scola} (we recall that on complex spaces this
peculiarity holds also if the Hamiltonian is time dependent but
quasi-stationary \cite{timdep}).

Let a Hermitian nonsingular quaternionic operator $\eta $ be
given. Then, the more general $\eta $-pseudoanti-Hermitian
quaternionic Hamiltonian $H$ can be written in the following
way:{\samepage
\begin{gather}
H=A\eta ,  \label{fact}
\end{gather}%
where $A^{\dagger }=-A$.}

In fact, let $H$ be given. Then, from the invertibility of $\eta $
the solution of equation~(\ref{fact}) can be immediately computed:
\begin{gather*}
A=H\eta ^{-1}.
\end{gather*}%
Now, from the $\eta $-pseudoanti-Hermiticity of $H$ we get
\begin{gather*}
A^{\dagger }=\eta ^{-1}H^{\dagger }=-\eta ^{-1}\eta H\eta
^{-1}=-H\eta ^{-1}=-A.
\end{gather*}

Let us consider now the time evolution of a pure state. Whenever
the
Hamiltonian $H$ is $\eta $-pseudoanti-Hermitian, the evolution operator $%
V(t)=e^{-Ht}$:
\begin{gather}
|\psi (t)\rangle =V(t)|\psi (0)\rangle  \label{evolutestate}
\end{gather}%
is no longer unitary, but $\eta $-unitary, i.e., it satisf\/ies%
\begin{gather}
V^{\dagger }\eta V=\eta .  \label{etaunitary}
\end{gather}%

From equations~(\ref{evolutestate}) and (\ref{etaunitary}), by
easy calculations, we obtain
\begin{gather*}
|\psi (t)\rangle \langle \psi (t)|\eta =V(t)|\psi (0)\rangle
\langle \psi (0)|V^{\dagger }(t)\eta =V(t)|\psi (0)\rangle \langle
\psi (0)|\eta
V^{-1}(t),  %\label{etanormconserv}
\end{gather*}%
or, equivalently,
\begin{gather}
\rho \left( t\right) \eta =\tilde{\rho}\left( t\right) =V(t)\tilde{\rho}%
(0)V(t)^{-1},  \label{conserbis}
\end{gather}%
from which the conservation of the $\eta $-pseudo-norm immediately
follows:
\begin{gather*}
{\rm Re}\,\mathrm{Tr}\tilde{\rho}(t)={\rm
Re}\,\mathrm{Tr}\,\tilde{\rho}(0).
\end{gather*}

More generally, the time evolution of $\rho (t)$ is described by
the equation
\begin{gather*}
\frac{d}{dt}\rho \left( t\right) =-(H\rho -\rho H^{\dagger }),
%\label{lioumod}
\end{gather*}%
($\hslash =1)$, whereas the time evolution of a generalized
density matrix (if $\eta $ is time independent) is described by
the usual Liouville--von Neumann equation:
\begin{gather*}
\frac{d}{dt}\tilde{\rho}\left( t\right) =-[H,\tilde{\rho}].
%\label{liouville}
\end{gather*}

\section{Complex projection of quasianti-Hermitian quaternionic\\ Hamiltonian
dynamics}

In this section, we restrict ourselves to consider the space of
quasi-Hermitian density matrices, that is the subclass of $\eta
$-pseudo-Hermitian density matrices where $\eta =B^{\dagger }B$ \
for some nonsingular bounded operator $B$.

An important property of such generalized density matrices is that
they are
positive def\/inite; indeed, putting $\eta =B^{\dagger }B$ into equation~(\ref{rho}), from the positivity of $\rho $ we immediately obtain $B\tilde{\rho}%
B^{-1}=B\rho B^{\dagger }\geq 0$~\cite{Zhang}.

Then, in this case a new positive def\/inite inner product can be
introduced in the Hilbert space where all the usual requirements
for a proper quantum mechanical interpretation can be maintained
\cite{altern,geyer,indefinite,mosta}.

In order to discuss the complex projection of $\eta
$-quasi-Hermitian quaternionic density matrices, we preliminarily
recall the following two lemmas~\cite{altern}:

\begin{lemma} \label{lemma1} For any Hermitian positive definite $\eta $, the (right) quaternionic Hilbert space $\mathbb{Q}^{n}$
endowed with the scalar product $(\cdot ,\cdot )_{\eta }=(\cdot
,\eta \cdot )$ is a Hilbert space $\mathbb{Q}_{\eta }^{n}$.
\end{lemma}

As a consequence of Lemma~\ref{lemma1} we can state that any $\eta
$-quasi-Hermitian operator in $\mathbb{Q}^{n}$, is Hermitian in
$\mathbb{Q}_{\eta }^{n}$. Moreover, as a particular case of the
lemma in~\cite{altern}, the following statement holds:

\begin{lemma}\label{lemma2} Let $\{\tilde{\rho}\}$ be an irreducible
set of $\eta $-quasi-Hermitian quaternionic operators on the
(right) quaternionic Hilbert space $\mathbb{Q}_{\eta }^{n}$
endowed with the scalar product $(\cdot ,\cdot )_{\eta }=(\cdot
,\eta \cdot )$. Suppose that $\{\tilde{\rho}\}$ cannot be
represented in
any basis by complex or real operators. Then, the commutant of $\{\tilde{%
\rho}\}$ is composed of the operators $T=h\mathbf{1}$ (where,
$h\in \mathbb{R}$  and $\mathbf{1}$ is the identity operator).
\end{lemma}

Now, we are able to prove the following proposition, which
characterizes the subclass of quaternionic $\eta $-quasi-Hermitian
density matrices, $\tilde{\rho}=\rho \eta =\tilde{\rho}_{\alpha
}+j\tilde{\rho}_{\beta }$, admitting $\eta $-quasi-Hermitian
complex projection density matrices $\tilde{\rho}_{\alpha }$.

\begin{proposition}\label{proposition1}
The complex projection $\tilde{\rho}%
_{\alpha }$ of a $\eta $-quasi-Hermitian quaternionic matrix
$\tilde{\rho}=$ $\tilde{\rho}_{\alpha }+j\tilde{\rho}_{\beta }$ is
$\eta $-quasi-Hermitian if and only if the entries of $\eta $ are
complex.
\end{proposition}

\begin{proof}
By imposing the $\eta $-quasi-Hermiticity of the complex
projection of $\tilde{\rho}$, from
equation~(\ref{complexProjection}) we get:
\begin{gather*}
\eta P[\tilde{\rho}]\eta ^{-1} =\tfrac{1}{2}(\tilde{\rho}^{\dagger
}-\eta i\eta ^{-1}\tilde{\rho}^{\dagger }\eta i\eta ^{-1})=
P^{\dagger }[\tilde{\rho}] =P[\tilde{\rho}^{\dagger }]=\tfrac{1}{2}(\tilde{%
\rho}^{\dagger }-i\tilde{\rho}^{\dagger }i),
\end{gather*}
hence,
\begin{gather}
(\eta ^{-1}i\eta i)\tilde{\rho}=\tilde{\rho}(\eta ^{-1}i\eta i).
\label{etacomm}
\end{gather}%
Now, equation~(\ref{etacomm}) must hold for any quaternionic $\eta $%
-quasi-Hermitian density matrix $\tilde{\rho}$, and the set $\{\tilde{\rho}%
\} $ is obviously a quaternionic irreducible set of $\eta
$-quasi-Hermitian matrices which cannot be represented in any
basis by real or complex operators. Then, Lemma \ref{lemma2}
implies
\begin{gather}
i\eta i=h\eta \qquad (h\in \mathbb{R}).  \label{etaant}
\end{gather}%
By using the cyclic property of the real part of the trace and
equation (\ref{etaant}) we get
\begin{gather*}
{\rm Re}\,\mathrm{Tr}(i\eta i)={\rm Re}\,\mathrm{Tr}(i^{2}\eta
)=-{\rm Re}\, \mathrm{Tr}(\eta )={\rm Re}\,\mathrm{Tr}(h\eta
)=h{\rm Re}\,\mathrm{Tr}(\eta ).
\end{gather*}%
Now, the positivity of $\eta $ implies, ${\rm
Re}\,\mathrm{Tr}(\eta )>0$, hence $h=-1$ and
equation~(\ref{etaant}) becomes
\begin{gather*}
\eta i=i\eta . \tag*{\qed} %\label{compl}
\end{gather*}\renewcommand{\qed}{}
\end{proof}

An important consequence of Proposition \ref{proposition1} is
that, if the complex projection $\tilde{\rho}_{\alpha }$ of
$\tilde{\rho}$ is $\eta $-quasi-Hermitian, then,
$\tilde{\rho}_{\alpha }$ is positive def\/inite; indeed,
$\tilde{\rho}_{\alpha }=\rho _{\alpha }\eta $ where $\eta
=B^{\dagger }B$, from the positivity of $\rho _{\alpha }$ we
immediately get $B\tilde{\rho}_{\alpha }B^{-1}=B\rho _{\alpha
}B^{\dagger }\geq 0$. Hence, from equation~(\ref{ex}) we can
conclude that if complex $\eta $-quasi-Hermitian operators $Q$ are
taken into account, the expectation value of $Q$ on
$\tilde{\rho}=$ $\tilde{\rho}_{\alpha }+j\tilde{\rho}_{\beta }$ or
on its complex projection $\tilde{\rho}_{\alpha }$ are the same.
This observation enables us to merge quasi-Hermitian complex
quantum mechanics in the more general framework of quasi-Hermitian
quaternionic quantum mechanics, without modifying any theoretical
prediction as long as complex observables are taken into account.

Now, we are able to discuss the complex projection of the $\eta
$-quasianti-Hermitian quaternionic Hamiltonian dynamics in the
space of $\eta $-quasi-Hermitian quaternionic density matrices.
Let be given the complex positive def\/inite operator $\eta $.
Then, the most general $\eta $-quasianti-Hermitian quaternionic
Hamiltonians $H$ reads (see equation~(\ref{fact}))
\begin{gather}
H=H_{\alpha }+jH_{\beta }=(A_{\alpha }+jA_{\beta })\eta ,
\label{GenFormH}
\end{gather}
where $A_{\alpha }^{\dagger }=-A_{\alpha }$ and $A_{\beta
}^{T}=A_{\beta }$. In other words, any $H$ is obtained by adding
to a complex $\eta $-quasianti-Hermitian Hamiltonian $A_{\alpha
}\eta $ a purely quaternionic $\eta $-quasianti-Hermitian
potential term $jA_{\beta }\eta $.

Moreover, putting $\eta =B^{\dagger }B$ into equation
(\ref{GenFormH}) we get
\begin{gather*}
H^{\prime }=BHB^{-1}=-(BHB^{-1})^{\dagger }=-H^{\prime \dagger },
%\label{Hermitian}
\end{gather*}
i.e., if we perform in the Hilbert space $\mathbb{Q}^{n}$ the
linear transformation induced by $B$, the Hamiltonian $H^{\prime
}$ is anti-Hermitian.

In particular, as said in Section~2, a standard master equation
will hold for the complex projection (see
equation~(\ref{gencompro})) of the dynamics generated by
$H^{\prime }$
\begin{gather}
\frac{d}{dt}\rho _{\alpha }^{\prime }=L[\rho _{\alpha }^{\prime
}]=-[H_{\alpha }^{\prime },\rho _{\alpha }^{\prime }]+H_{\beta
}^{\prime \ast }\rho _{\beta }^{\prime }-\rho _{\beta }^{\prime
\ast }H_{\beta }^{\prime },  \label{mastereq}
\end{gather}%
where $\rho _{\alpha }^{\prime }$ and $\rho _{\beta }^{\prime }$
are the complex projection and the purely quaternionic terms,
respectively, of the quaternionic (Hermitian) density matrix $\rho
^{\prime }=\rho _{\alpha }^{\prime }+j\rho _{\beta }^{\prime
}=B\rho \eta B^{-1}=B\rho B^{\dagger }$.

The dynamics ruled by equation (\ref{mastereq}) is a one-parameter
semigroup dynamics in the space of complex (Hermitian) density
matrices~\cite{Asor}, and we can identify~\cite{Kossa3,lin}
\begin{gather*}
L[\rho _{\alpha }^{\prime }]=-[H_{\alpha }^{\prime },\rho _{\alpha
}^{\prime }]+\sum_{r,s=1}^{n^{2}-1}C_{rs}\left(F_{r}^{\prime }\rho
_{\alpha }^{\prime }F_{s}^{\prime \dagger
}-\tfrac{1}{2}\{F_{r}^{\prime \dagger }F_{s}^{\prime
},\rho _{\alpha }^{\prime }\}\right), % \label{kossagenerator}
\end{gather*}%
where $F_{r}^{\prime }$ are $n^{2}-1$ traceless square matrices,
which form with the normalized identity $F_{0}^{\prime
}=I_{n}/\sqrt{n}$ an orthonormal set, i.e.,
$\mathrm{Tr}(F_{r}^{\prime \dagger }F_{s}^{\prime })=\delta _{rs}
$, while $[C_{rs}]$ is a Hermitian matrix.

Then, coming back by means of $B^{-1}$: $\left| \psi ^{\prime
}\right\rangle \rightarrow B$ $^{-1}\left| \psi ^{\prime
}\right\rangle $, we obtain an equation of the
Lindblad--Kossakowski type which describes the most general
time evolution of the generalized complex projection density matrix $\tilde{%
\rho}_{\alpha }=\rho _{\alpha }\eta $. In order to write down it
explicitly, let us recall that a density matrix transforms as
follows: $\rho _{\alpha }^{\prime }\rightarrow B^{-1}\rho _{\alpha
}^{\prime }(B^{-1})^{\dagger }=\rho _{\alpha }$, so that
$\tilde{\rho}_{\alpha }=\rho _{\alpha }\eta =B^{-1}\rho _{\alpha
}^{\prime }(B^{-1})^{\dagger }\left( B^{\dagger }B\right)
=B^{-1}\rho _{\alpha }^{\prime }B$. Moreover, $B^{-1}F_{r}^{\prime
}B=F_{r}$ and $B^{-1}F_{s}^{\prime \dagger
}B=B^{-1}(BF_{s}B^{-1})^{\dagger }B=$ $F_{s}^{\ddagger }$
according to (\ref{adj}), so that, f\/inally \cite{bologna}:
\begin{gather}
\frac{d}{dt}\tilde{\rho}_{\alpha }=L[\tilde{\rho}_{\alpha }]=-[H_{\alpha },%
\tilde{\rho}_{\alpha }]+\sum_{r,s=1}^{n^{2}-1}C_{rs}\left(F_{r}\tilde{\rho}%
_{\alpha }F_{s}^{\ddagger }-\tfrac{1}{2}\{F_{r}^{\ddagger }F_{s},\tilde{\rho}%
_{\alpha }\}\right).  \label{kossaquasihermitian}
\end{gather}%
Note that trivially $\mathrm{Tr}(F_{r}^{\ddagger }F_{s})=\delta
_{rs}$ and the dissipative term
\begin{gather}
D[\tilde{\rho}_{\alpha }]=\sum_{r,s=1}^{n^{2}-1}C_{rs}\left(F_{r}\tilde{\rho}%
_{\alpha }F_{s}^{\ddagger }-\tfrac{1}{2}\{F_{r}^{\ddagger }F_{s},\tilde{\rho}%
_{\alpha }\}\right)=H_{\beta }^{\ast }\tilde{\rho}_{\beta
}-\tilde{\rho}_{\beta }^{\ast }H_{\beta },  \label{dissipative}
\end{gather}%
is quasi-Hermitian:
\begin{gather*}
\eta D[\tilde{\rho}_{\alpha }]\eta ^{-1}=D[\tilde{\rho}_{\alpha
}]^{\dagger }.
\end{gather*}

Finally, the complex projection of quaternionic anti-Hermitian
Hamiltonian dynamics can be immediately obtained as a very
particular case of this more general setting, putting $\eta
=\mathbf{1}$ into equations~(\ref{kossaquasihermitian}),
(\ref{dissipative}).

\section[A spin-$\frac{1}{2}$ system in a constant quasianti-Hermitian
quaternionic potential]{A spin-$\boldsymbol{\frac{1}{2}}$ system in a constant quasianti-Hermitian\\
quaternionic potential}

We will now consider a two-level quantum system with a
quasianti-Hermitian quaternionic $H=H_{\alpha }+jH_{\beta }$.

We denote by $H_{\alpha }$ the free complex anti-Hermitian
Hamiltonian describing a spin half particle in a constant magnetic
f\/ield,
\begin{gather*}
H_{\alpha }=\frac{\omega }{2}\left(
\begin{array}{cc}
i & 0 \\
0 & -i%
\end{array}%
\right) ,  %\label{complexFreeHamiltonian}
\end{gather*}%
and by $jH_{\beta }$\ a purely quasianti-Hermitian quaternionic
constant potential,
\begin{gather}
jH_{\beta }=\left(
\begin{array}{cc}
0 & j\frac{v}{x} \\
jvx & 0%
\end{array}%
\right) \qquad (v,x\in \mathbb{R}\backslash \{0\}).
\label{quaternionicPotential}
\end{gather}%
Note that $H_{\alpha }$ and $jH_{\beta }$ are $\eta
$-quasianti-Hermitian, where
\begin{gather*}
\eta =\left(
\begin{array}{cc}
x^{2} & 0 \\
0 & 1%
\end{array}%
\right) .  %\label{etaes}
\end{gather*}

The eigenvalues and the corresponding biorthonormal eigenbasis of
the quaternionic Hamiltonian $H$ are given by \cite{scola}
\begin{gather*}
iE_{\pm }=i\left( \frac{\omega }{2}\pm v\right)  %\label{eigenvalues}
\end{gather*}%
and
\begin{gather*}
|\psi _{\pm }\rangle =\left(
\begin{array}{c}
\pm \frac{i}{x} \\
j%
\end{array}%
\right) \frac{1}{\sqrt{2}},\qquad |\phi _{\pm }\rangle =\left(
\begin{array}{c}
\pm xi \\
j%
\end{array}%
\right) \frac{1}{\sqrt{2}}.  %\label{Heigenvectors}
\end{gather*}

The $\eta $-unitary evolution operator reads
\begin{gather*}
V(t) =e^{-Ht}=|\psi _{+}\rangle e^{-iE_{+}t}\langle \phi
_{+}|+|\psi
_{-}\rangle e^{-iE_{-}t}\langle \phi _{-}|  \\ %\label{evolutionoperator} \\
\phantom{V(t)}{}=\frac{1}{2}\left(
\begin{array}{cc}
e^{-iE_{+}t}+e^{-iE_{-}t} & \frac{1}{x}(e^{-iE_{-}t}-e^{-iE_{+}t})k \vspace{1mm}\\
x(e^{iE_{+}t}-e^{iE_{-}t})k & e^{iE_{+}t}+e^{iE_{-}t}%
\end{array}%
\right) .  \notag
\end{gather*}

Let us consider a $\eta $-quasi-Hermitian complex pure initial
state:
\begin{gather*}
\tilde{\rho}(0)=\left(
\begin{array}{cc}
0 & 0 \\
0 & 1%
\end{array}%
\right) ,  %\label{initial}
\end{gather*}%
then, (see equations (\ref{conserbis}) and (\ref{etaunitary}))
\begin{gather}\label{final}
\tilde{\rho}(t)
=V(t)\tilde{\rho}(0)V(t)^{-1}=V(t)\tilde{\rho}(0)\eta
^{-1}V^{\dagger }(t)\eta = \frac{1}{2}\left(
\begin{array}{cc}
1-\cos (2vt) & -\frac{j}{x}\sin (2vt) \\
jx\sin (2vt) & 1+\cos (2vt)%
\end{array}%
\right) .
\end{gather}%
The complex projection $\tilde{\rho}_{\alpha }(t)$ of
$\tilde{\rho}(t)$ assumes the diagonal form,
\begin{gather*}
\tilde{\rho}_{\alpha }(t)=\frac{1}{2}\left(
\begin{array}{cc}
1-\cos (2vt) & 0 \\
0 & 1+\cos (2vt)%
\end{array}%
\right) .
\end{gather*}

The one-parameter semigroup generator associated with the complex
projection of the quaternionic $\eta $-unitary dynamics given in
equation (\ref{final}) can
be immediately computed (see equations (\ref{kossaquasihermitian}), (\ref%
{dissipative})):
\begin{gather}
L[\tilde{\rho}_{\alpha }(t)]=-[H_{\alpha },\tilde{\rho}_{\alpha
}]+H_{\beta }^{\ast }\tilde{\rho}_{\beta }-\tilde{\rho}_{\beta
}^{\ast }H_{\beta }=\left(
\begin{array}{cc}
v\sin (2vt) & 0 \\
0 & -v\sin (2vt)%
\end{array}%
\right) .  \notag
\end{gather}

Let us consider the spin observable, which is associated in QQM
with a triple of complex operators. The expectation value of the
$z$-component:
\begin{gather*}
s_{z}=\frac{1}{2}\left(
\begin{array}{cc}
1 & 0 \\
0 & -1%
\end{array}%
\right) , % \label{spinobsevable}
\end{gather*}%
when the system is in the quasi-Hermitian quaternionic pure state
(\ref{final}), is given by (note that $s_{z}$ is $\eta
$-quasi-Hermitian contrarily to $s_{x}$ and $s_{y}$)
\begin{gather}
\langle s_{z}\rangle ={\rm Re}\,\mathrm{Tr}\left(
s_{z}\tilde{\rho}(t)\right)
=\mathrm{Tr}\left( s_{z}\tilde{\rho}_{\alpha }(t)\right) =\frac{\cos (2vt)}{2%
}.  \label{expectspiobservable}
\end{gather}

By a simple calculation the (positive def\/inite) energy $\eta
$-quasi-Hermitian observable $|H|$ reads
\begin{gather*}
|H|=|\psi _{+}\rangle E_{+}\langle \phi _{+}|+|\psi _{-}\rangle
E_{-}\langle \phi _{-}|=\left(
\begin{array}{cc}
\frac{\omega }{2} & -k\frac{v}{x} \vspace{1mm}\\
kxv & \frac{\omega }{2}%
\end{array}%
\right) ,  %\label{energyobservable}
\end{gather*}%
and its expectation value is given by
\begin{gather}
\langle |H|\rangle ={\rm Re}\,\mathrm{Tr}\left( |H|\tilde{\rho}(t)\right) =%
\mathrm{Tr}\left( |H_{\alpha }|\tilde{\rho}_{\alpha }(t)\right) =\frac{%
\omega }{2}={\rm Re}\,\mathrm{Tr}\left( |H|\tilde{\rho}(0)\right)
. \label{energyexpectation}
\end{gather}

This example may have interesting physical applications because
the quaternionic poten\-tial~$jH_{\beta }$ in equation
(\ref{quaternionicPotential}) strongly af\/fects the spin values
(see equation~(\ref{expectspiobservable})) while the system energy
is unchanged (see equation~(\ref{energyexpectation})).

\pdfbookmark[1]{References}{ref}
\LastPageEnding


\begin{thebibliography}{99}

\footnotesize\itemsep=0pt

\bibitem{bir} Birkhof\/f G., von Neumann J., The logic of quantum mechanics,
\textit{Ann. Math.} \textbf{37} (1936), 823--843.

\bibitem{fin} Finkelstein D., Jauch J.M., Schiminovich  S., Speiser D.,
Foundations of quaternion quantum mechanics, \textit{J.~Math. Phys.} \textbf{3} (1962), 207--220.\\
 Finkelstein D., Jauch J.M., Schiminovich  S., Speiser D.,
 Principle of general $Q$-covariance, \textit{J. Math. Phys.}  \textbf{4} (1963),
788--796.\\
 Finkelstein D., Jauch  J.M., Speiser D., Quaternionic
representations of compact groups, \textit{J. Math. Phys.}
\textbf{4} (1963), 136--140.

\bibitem{Adler} Adler S.L., Quaternionic quantum mechanics and quantum
f\/ields, Oxford University Press, New York, 1995.

\bibitem{Kossa1} Kossakowski A., Remarks on positive maps of
f\/inite-dimensional simple Jordan algebras, \textit{Rep. Math.
Phys.} \textbf{46} (2000), 393--397.

\bibitem{compent} Scolarici G., Solombrino L., Complex entanglement and
quaternionic separability, in The Foundations of Quantum
Mechanics: Historical Analysis and Open Questions (Cesena, 2004),
Editors C.~Garola, A.~Rossi and S.~Sozzo, World Scientif\/ic,
Singapore, 2006, 301--310.

\bibitem{Asor} Asorey  M., Scolarici G., Complex positive maps and
quaternionic unitary evolution, \textit{J. Phys. A: Math. Gen.}
\textbf{39} (2006), 9727--9741.

\bibitem{gallipoli} Asorey M., Scolarici G., Solombrino L., Complex
projections of completely positive quaternionic maps,
\textit{Theoret. and Math. Phys.} \textbf{151} (2007), 735--743.

\bibitem{ASS} Asorey M., Scolarici G., Solombrino L., The complex
projection of unitary dynamics of quaternionic pure states,
\textit{Phys. Rev. A} \textbf{76} (2007), 012111--012117.

\bibitem{proc} Proceedings of the Ist, IInd, IIIrd, IVth and Vth
International Workshops on ``Pseudo-Hermitian Hamiltonians in
Quantum Physics'' in \textit{Czech. J. Phys.} \textbf{54} (2004),
no.~1, no.~10, \textit{Czech. J. Phys.} \textbf{55} (2005),
no.~9, \textit{J.~Phys.~A: Math. Gen.} \textbf{39} (2006),
no.~32, and \textit{Czech. J. Phys.} \textbf{56} (2006), no.~9,
respectively.

\bibitem{scola} Scolarici G., Pseudoanti-Hermitian operators in quaternionic
quantum mechanics, \textit{J. Phys. A: Math. Gen.} \textbf{35}
(2002), 7493--7505.

\bibitem{altern} Blasi A., Scolarici  G., Solombrino L., Alternative
descriptions in quaternionic quantum mechanics, \textit{J.~Math.
Phys.} \textbf{46} (2005), 42104--42111,
\href{http://arxiv.org/abs/quant-ph/0407158}{quant-ph/0407158}.

\bibitem{bologna} Scolarici  G., Solombrino L., Time evolution of
non-Hermitian quantum systems and generalized master equations,
\textit{Czech. J. Phys.} \textbf{56} (2006), 935--941.

\bibitem{obse} Mostafazadeh A., Batal A., Physical aspects of
pseudo-Hermitian and PT-symmetric quantum mechanics, \textit{J.
Phys. A: Math. Gen.} \textbf{37} (2004), 11645--11680,
\href{http://arxiv.org/abs/quant-ph/0408132}{quant-ph/0408132}.

\bibitem{timdep} Mostafazadeh A., Time-dependent pseudo-Hermitian
Hamiltonians def\/ining a unitary quantum system and uniqueness of
the metric operator, \textit{Phys. Lett. B} \textbf{650} (2007),
208--212, \href{http://arxiv.org/abs/0706.1872}{arXiv:0706.1872}.

\bibitem{Zhang} Zhang F., Quaternions and matrices of quaternions, \textit{Linenar Algebra Appl.} \textbf{251} (1997), 21--57.

\bibitem{geyer} Sholtz F.G., Geyer H.B., Hahne F.J.W.,
Quasi-Hermitian operators in quantum mechanics and the variational
principle, \textit{Ann. Phys.} \textbf{213} (1992), 74--101.

\bibitem{indefinite} Blasi A., Scolarici G., Solombrino L.,
Pseudo-Hermitian Hamiltonians, indef\/inite inner product spaces
and their symmetries, \textit{J. Phys. A: Math. Gen.} \textbf{37}
(2004), 4335--4351,
\href{http://arxiv.org/abs/quant-ph/0310106}{quant-ph/0310106}.

\bibitem{mosta} Mostafazadeh A., Exact PT-symmetry is equivalent to
Hermiticity, \textit{J. Phys. A: Math. Gen.} \textbf{36} (2003),
7081--7092,
\href{http://arxiv.org/abs/quant-ph/0304080}{quant-ph/0304080}.

\bibitem{Kossa3} Gorini V., Kossakowski A., Sudarshan E.C.G.,
Completely positive dynamical semigroups and $N$-level systems,
\textit{J.~Math. Phys.} \textbf{17} (1976), 821--825.

\bibitem{lin} Lindblad G., On the generators of quantum dynamical
semigroups, \textit{Comm. Math. Phys.} \textbf{48} (1976),
119--130.
\end{thebibliography}
\end{document}